\documentstyle[12pt]{article}
\title{Single heavy lepton production in high energy
electron-positron collisions}
\author{F. M. L. Almeida Jr, J. H. Lopes, J. A. Martins Sim\~oes,\\
P. P. Queiroz Filho, A. J. Ramalho\\
Instituto de F\'{\i}sica\\
Universidade Federal do Rio de Janeiro\\
Ilha do Fund\~ao, Rio de Janeiro\\
21945-970, RJ, Brazil}
\date{}
\begin{document}

\maketitle

\begin{abstract}
The production and decay of exotic leptons are discussed in the
context of the vector singlet, vector doublet and fer\-mion mirror-fer\-mion
models, at c.m. energies $\sqrt{s}=190GeV$ (LEP II) and $\sqrt{s}=500GeV$
(NLC).   The model dependence of total cross sections and kinematic
distributions at these energies is shown to be weak for these three
extended models, indicating that it would be difficult to establish which
of them is best fit to describe the underlying new physics, should exotic
leptons be detected.   We suggest that, in order to resolve this difficulty,
one should measure the angular distributions of single exotic leptons in
longitudinally polarized $e^+e^-$ collisions.  
\end{abstract}
\newpage

\section*{I.Introduction}
There exist at present many theoretical models that could
generalize  the standard scenario: composite models, grand
unified theories, supersymmetry, technicolor, mirror fermions,
superstrings, etc.   All these models show a common feature - they
include new particles besides the ones presently known.   The next
generation of high energy physics experiments will decide which of
these models, if any, corresponds to physical reality.   If we take an
optimistic point of
view and suppose  that  at  HERA  and/or $e^+e^-$ energies up to 500 GeV some
"new physics" will be found, then the major question will be
the   theoretical  origin of  any  new  object ( hopefully) seen by
the experimentalists.  Since  extended  models  have a large number  of
undetermined   parameters -masses, mixing angles, etc.-  this  will
be a very difficult question to  answer.  In this paper we only deal with
the lepton sector. In previous publications \cite{Alm91,Alm93} we
compared  predictions  for exotic-lepton  production  and  decays  in
some of the  most popular extended models: vector  singlet
(VSM) \cite{Val},  vector  doublet (VDM)\cite{Hew89}  and fermion
mirror-fermion model
(FMFM) \cite{Csi94}. In ref.\cite{Alm93}  we explicitly showed that it
would be very difficult
to establish which of these three models might account for any eventual
exotic lepton that materializes in electron-proton collisions
at present energies.
\par
 Here we consider ${e}^+{e}^-$ collisions at  $\sqrt{s} = 190 GeV$
 (LEP II) and at a  future  collider (NLC) with $\sqrt{s} = 500 GeV$.
  New  heavy lepton pair production has been extensively studied in the
literature \cite{Buc93} and is kinematically bounded by ${M}_L \leq
\sqrt{s}/2$.
All models predict mixing between light and heavy leptons and allow
single heavy lepton production \cite{Gan89} with masses up to $\sqrt{s}$. This
was investigated in ref.\cite{Alm91} for  LEP I energies ,
and we extend our previous analysis here for higher energies.  In  order  to
reduce the number of theoretical inputs, we have assumed that at these
energies  new leptons interact only via the ordinary gauge vector bosons
$\gamma$, ${W}^{\pm}$ and ${Z}^0$.
Other gauge bosons are predicted by many extended models, but the
resulting new interactions produce small effects at the current energy scale.
Our  results are  organized as follows: in section II we  review  the
extended models, their  couplings,  bounds and decay properties; in section
III we consider unpolarized electron-positron  collisions; in section IV  we
consider the production of exotic leptons with longitudinally polarized
electron and positron beams, and in section V we briefly state our conclusions.

\section*{II.Models for new leptons}
The general charged and neutral interactions of the new
leptons $(L^0, L^{\pm})$ mixed with the usual ones and the standard
W and Z bosons can be written as
\begin{equation}
{L}_{CC} = {-g\over {2\sqrt {2}}}\{{\bar L}^0 {\gamma}^{\mu}
({a}_1-{b}_1{\gamma}^5) {e}^-  + {\bar {\nu}} {\gamma}^{\mu}
({a}_2-{b}_2{\gamma}^5) {L}^- \} {W}_{\mu}
\end{equation}
and
\begin{equation}
{L}_{NC} = {-g\over {4{\cos {\theta}_W}}}{\sum_{i,j}}
\{{\bar f}_i {\gamma}^{\mu}({v}_{ij}-{a}_{ij}{\gamma}^5){f}_j\}{Z}_{\mu}
\end{equation}

        The coupling constants in eqs.(1) and (2) are given in table I of
ref.\cite{Alm91}
for each of the three extended models.  Upper bounds on the mixing angles
were obtained from phenomenological analysis \cite{Lan92} and lead to
${sin}^2{\theta}_i {< \atop \sim} 0.01-0.03$.
The large number of mixing parameters imply a large number of couplings in
each model. We have taken into account the limiting situation of equal
angles which give $V \pm A$, V or A couplings.  In our estimates we have
taken all mixing angles to have the value ${\theta}_i = 0.1$.
 This value is not a limiting  bound and can be easily scaled.
Experimental bounds on heavy lepton masses set a lower bound around
the  Z  mass \cite{L3C92}.  In  the  mass  region  ${M}_L > {M}_Z$  the
dominant  decay  is
given by $L \longrightarrow l B$, where B denotes either the W or the Z boson
and $l$ an ordinary lepton.
For the decay $L({p}_L) \longrightarrow l({p}_1) B({p}_2)$ and a
general vertex of the form $-ie {\gamma}^{\mu} (C_v-C_a {\gamma}^5)$
the width is equal to
\begin{equation}
\Gamma (L \longrightarrow Bl) ={\displaystyle
{1\over 4} \alpha M_L (C_v^2+C_a^2)(y-\frac{3}{y} + \frac{2}{y^2})},
\end{equation}
where
$y ={\displaystyle (\frac{M_L}{M_B})^2}$, whereas the differential decay
rate is
given by
\begin{eqnarray}
d\Gamma &=& {1\over {E}_L}{\alpha \over \pi} \{ ({C_v}^2+{C_a}^2)
[{p}_1.{p}_L +{2\over {{M}_2}^2}{p}_1.{p}_2{p}_2.{p}_L]-
3(C_v^2-C_a^2){M}_1{M}_L\}\nonumber\\
&\ &\times {\delta}^{(4)}({p}_L-{p}_1-{p}_2)
{{d}^3{p}_1 \over {2{E}_1}}{{d}^3{p}_2 \over {2{E}_2}}
\end{eqnarray}
\par
The electromagnetic decay $L \longrightarrow l \gamma$ is strongly
suppressed for the models we are considering here, since they can occur only
via one-loop processes.  For an exotic charged lepton the dominant
decay mode is ${L}^- \longrightarrow {{\nu}_l} {W}^-$.
The most favored final state is ${L}^- \longrightarrow {\nu}_l + 2 jets$,
with a branching ratio of $67 \%$. The clean leptonic final states
${\nu}_l ({e}^- \bar {\nu}_e)$ and ${\nu}_l ({\mu}^- \bar {\nu}_{\mu})$
have a branching ratio of $11\%$ each. For a neutral heavy lepton the main
decay
mode is ${L}^0 \longrightarrow {l}^- {W}^+$, and the values of the branching
ratios for each mode are similar to the corresponding ratios in the case of
charged exotic leptons.   A final state with a real Z boson is also
possible but suppressed relative to the W boson case.  We draw attention to an
important point concerning this two-body decay.  In the familiar three-body
fermion decay, the differential decay rates show a linear dependence
on the vector and axial couplings \cite{Alm93}.  This simple fact leads to
a different
energy spectrum for the final fermions in the V+A, V-A and V, A
cases.
In the two-body decay one can see from eq.(4) that this is no longer
true.   This means that no model dependence can be established from the
decay spectrum only.   Hence one  must look for differences in the basic
production mechanism.
\par

\section*{III.Unpolarized electron-positron collisions}

The general expression for the angular distribution of single
exotic leptons in unpolarized ${e}^+{e}^-$ collisions was given in the
appendix of ref.\cite{Alm91}.  The dependence of the total cross section on
the mass
of the heavy charged lepton is displayed in Figs. 1-2, for c.m. energies
$\sqrt{s} = 190 GeV$ and $\sqrt{s} = 500 GeV$.   The analogous curves for
neutral lepton production are given in Figs. 3-4.   In ref.\cite{Alm91} it
was shown that
at $\sqrt{s} = {M}_Z$ the angular distributions of the primary lepton
(electron
or reconstructed neutrino) could be used to distinguish the three extended
model, both in the $L^\pm e^\mp$ and $L^0 \bar\nu_e$
production modes.   In heavy-lepton production around the Z mass, the
s-channel contributions clearly dominate over those coming from t-channel
and interference terms.   The most recent experimental searches, however,
place the exotic-lepton mass most likely above the Z mass.
The s-channel contributions to the unpolarized angular spectra are suppressed
with respect to the t-channel contributions at LEP II, and even more
so at NLC energies.   Explicit calculations show
that at LEP II, and especially at NLC energies, the power of the
unpolarized angular
distributions of the exotic leptons to differentiate one nonstandard model
from another is rather limited, as depicted in Figs. 5-6.   These two graphs
correspond to the angular spectra of heavy neutral leptons, but the same
pattern is reproduced for exotic charged leptons.

\section*{IV.Polarized electron-positron collisions}
Polarized cross sections can often provide important
information which otherwise would be concealed in their unpolarized
counterparts.   In this section we argue that the angular
distributions of single exotic leptons, arising from the collisions
of longitudinally polarized ${e}^+{e}^-$ beams at LEP II or NLC energies, are
quite sensitive to the structure of the specific extended models
under study.   As discussed earlier, this model dependence is
negligible in the unpolarized distributions, both at LEP II and NLC
energies.   Expressions for the definite-helicity angular distributions
${d{\sigma}^{({h}_1,{h}_2)}/{d(cos{\theta})}}$ can be easily derived from
the definite-helicity invariant differential cross section
$d{\sigma}^{({h}_1,{h}_2)}/{dt}$
given in the appendix, where ${h}_1$ and ${h}_2$
stand for the helicities of the incoming electrons and positrons
respectively, and $\theta$ is the angle of the exotic-lepton three-momentum
with respect to the direction of the  electron beam.   Our numerical
calculations of spin-averaged cross sections and angular distributions are
consistent with those in refs.\cite{Alm91} and \cite{Maa92}.   In order to
reduce normalization
uncertainties, we computed the angular distributions
${\displaystyle {\cal A}^{(h_1,h_2)}(cos{\theta}) \equiv
{1\over{\sigma^{(h_1,h_2)}}}
{d{\sigma}^{(h_1,h_2)}\over{d(cos{\theta})}}}$,
which are normalized to the corresponding definite-helicity total
cross sections.

\par
Figs. 7-8 show polarized angular distributions
concerning the production of charged heavy leptons at LEP II energy $\sqrt
{s} = 190 GeV$ and a lepton mass ${M}_L = 120 GeV$.   Fig.7 shows the angular
distribution ${\cal A}^{(-,-)}(cos{\theta})$ for the FMFM and the VSM.
Only the
t-channel contribution is nonvanishing, since helicity conservation
forbids that the reaction proceed via the s channel if ${h}_1 = {h}_2$.
Moreover, the t-channel contribution vanishes in the case of the VDM on
account of the V+A character of the nonstandard neutral weak current in
this model.   When ${h}_1 = {h}_2 = 1$ the angular distributions for the VDM
and the FMFM look very much like the one shown in Fig.7, but now it is the
angular distribution for the VSM that is equal to zero.   This can be
understood in terms of helicity conservation and the V-A nature of the
nonstandard neutral weak current in the VSM.   Fig.8 displays the distinct
behaviors of the curves for ${\cal A}^{(-,+)}(cos{\theta})$ according to the
three extended models.   Since in this case the angular distribution for the
VDM arises entirely from the s-channel contribution, it will be suppressed
at higher energies.   Considering the reverse helicity combination
${h}_1= -{h}_2=1$, we obtain the same pattern of Fig.8, but now the roles
of the VDM and VSM are interchanged.   It is worthwhile to point out that
the production rates of charged exotic leptons with polarized beams are
not particularly high at LEP II energies.   We estimate rates of the order
of 5 events per year at a conservative luminosity of $10^2pb^{-1}/year$.
At a $500GeV$ NLC with a yearly luminosity of $10fb^{-1}$, on the
other hand, the production rates are about $10^3$ events per year.

\par
Next we discuss the angular spectra in the production of neutral
heavy leptons.   First of all, the t-channel contributions to
${\cal A}^{(-,-)}(cos{\theta})$ were found to vanish in all three models.
In the VDM this comes about because the nonstandard charged weak current
is V+A if all the mixing angles are taken to have the same value.
This choice of mixing angles and the V-A standard coupling
$W^{-}e^{+}\bar{\nu}$ are responsible for the result in the FMFM.   On the
other hand, for any set of mixing angles, both the standard and
nonstandard charged currents are V-A in the VSM, accounting for a
vanishing t-channel contribution.   Fig.9 illustrates the angular
distribution in the FMFM and the VSM for the helicity combination
${h}_1 = -{h}_2 = -1$.   The spectrum for the VDM is not shown in the
picture because the corresponding production rate was found to be much
smaller in this model.   The curve for ${\cal A}^{(+,+)}(cos{\theta})$ in
the VDM
and FMFM is similar to the one shown in Fig.9.   Here the t-channel
contribution in the VSM is also equal to zero.   Finally, Fig.10 illustrates
the angular dependence of ${\cal A}^{(+,-)}(cos{\theta})$ in the framework
of the
three models.   All the distributions  arise exclusively from the s-channel
process, and are bound to be suppressed at higher energies.   To conclude
this section, it is useful to draw attention to the fact that the
production rates of exotic neutral lepton are expected to be comparatively
higher in the FMFM and VSM for ${h}_1 = -{h}_2 = -1$, as well as in the
VDM and FMFM for ${h}_1 = {h}_2 = 1$.   Taking all mixing angles equal to
$0.1$, we obtain rates of the order of $10^2$ events/year at LEP II and
$10^4$ events/year at NLC.

\section*{V.Conclusions}
The present and future $e^+e^-$ colliders offer the opportunity to
explore new energy scales and search for manifestations of nonstandard
physics, which could turn out to be the discovery of exotic leptons.
Assuming that exotic lepton signatures are observable at LEP II or NLC,
it is appropriate to ask how to decide which extended model correctly
accounts for this new physics.   At LEP I energies, the angular
distribution of single exotic leptons is known to be sensitive to the
structure  of the three extended models in which we are interested, namely,
the vector doublet, vector singlet and the fermion mirror-fermion models.
Current experimental results suggest, however, that the heavy-lepton mass
is above the Z mass.   We have shown that, at LEP II and NLC energies, as
the s-channel annihilation no longer dominates over the t-channel
exchange, the angular distribution of exotic leptons ceases to distinguish
the models.   For this energy regime, we have demonstrated the utility of
measuring angular distributions of single exotic leptons produced in the
collisions  of longitudinally polarized $e^+e^-$ beams, with the purpose
of testing the various theories which have been proposed to generalize the
standard model of electroweak interactions.   In order to reduce potential
systematic errors, it might be more expedient for the experimentalists to
measure helicity asymmetries.

\par
{\it Acknowledgments:} This work was partly supported by CNPq and FINEP.
\newpage

\section*{Figure captions}
\newcounter{ijk}
\begin{list}%
{Fig.\arabic{ijk}\ }{\usecounter{ijk}\setlength{\rightmargin}{\leftmargin}}

\item Total cross sections as functions of the exotic charged lepton mass
$M_{L^-}$, for LEP II energy $\sqrt{s} = 190GeV$, according to the VDM,
FMFM and VSM.

\item Same as Fig.1, but for $\sqrt{s} = 500GeV$.

\item Total cross sections as functions of the exotic neutral lepton mass
$M_{L^0}$, for LEP II energy $\sqrt{s} = 190GeV$, according to the VDM,
FMFM, and VSM.

\item Same as Fig.3, but for $\sqrt{s} = 500GeV$.

\item Normalized angular distributions of an exotic neutral lepton at
$\sqrt{s} = 190GeV$, for an input mass $M_{L^0}=120GeV$

\item Same as Fig.5, but for $\sqrt{s}=500GeV$ and $M_{L^0}=300GeV$.

\item Normalized angular distribution ${\cal A}^{(-,-)}(cos{\theta})$ of
an exot\-ic charged lep\-ton according to the FMFM and the VSM, at LEP II,
for an input mass $M_{L^-} = 120GeV$.

\item Normalized angular distribution ${\cal A}^{(-,+)}(cos{\theta})$ of an
exot\-ic charged lep\-ton according to the FMFM, VDM and VSM, at LEP II,
for an input mass $M_{L^-} = 120GeV$.

\item Normalized angular distribution ${\cal A}^{(-,+)}(cos{\theta})$ of an
exotic neutral lepton according to the FMFM, VDM and VSM, at LEP II, for
an input mass $M_{L^0} = 120GeV$.

\item Normalized angular distribution ${\cal A}^{(+,-)}(cos{\theta})$ of an
exotic neutral lepton according to the FMFM, VDM and VSM, at LEP II, for
an input mass $M_{L^0} = 120GeV$.
\end{list}
\newpage

\section*{Appendix}
The definite-helicity invariant differential cross section
$d{\sigma}^{({h}_1,{h}_2)}/{dt}$ for the production of single exotic
leptons, through the reactions\\
$e^+ + e^- \longrightarrow L^-(L^0) + e^+(\bar \nu),$
 is given by
\begin{eqnarray}
{\displaystyle {d\sigma \over dt}^{(h_1,h_2)} =
{1\over {16\pi s^2}}\vert{{{\cal M}^{({h}_1,{h}_2)}}\vert}^2},
\end{eqnarray}
where
\begin{eqnarray}
{\vert{{\cal M}^{(h_1,h_2)}\vert}}^2 & = &
{{8\pi^2\alpha^2(1-h_1h_2)}\over {D(s,M_Z,\Gamma_Z)}}\nonumber\\
&\  &\times
\lbrace(c_{2V}^2+c_{2A}^2)
[(3-h_1h_2)(g^2_{2V}+g^2_{2A})+4(h_2-h_1) g^{\ }_{2V}g^{\ }_{2A}]\nonumber\\
&\ &\hskip 1cm \times[s(s+2t-M_L^2)+2t(t-M_L^2)] \nonumber\\
&\ &\hskip 0.5cm +4c_{2V}c_{2A}[(3-h_1h_2)g^{\ }_{2V}g^{\ }_{2A}+
(h_2-h_1)(g^2_{2V}+g^2_{2A})]\nonumber\\
&\ &\hskip 1cm \times s(s+2t-M_L^2)\rbrace\nonumber\\
&+& {{32\pi^2\alpha^2}\over {D(t,M_B,\Gamma_B)}}\nonumber\\
&\ &\times \lbrace [(g^2_{1V}+g^2_{1A})(c^2_{1V}+c^2_{1A}+2(h_2-h_1)c^{\
}_{1V}c^{\ }_{1A})\nonumber\\
&\ &\hskip 2.5cm -4h_1h_2c^{\ }_{1V}c^{\ }_{1A}g^{\ }_{1V}g^{\ }_{1A}]
\nonumber\\
&\ &\hskip 1cm\times [2s(s+t-M^2_L)+t(t-M^2_L)]\nonumber\\
&\ &\hskip 0.5cm
 -h_1h_2[(c^2_{1V}+c^2_{1A})(g^2_{1V}+g^2_{1A}+2(h_2-h_1)g^{\ }_{1V}
g^{\ }_{1A})\nonumber\\
&\ &\hskip 2.5cm  -4h_1h_2c^{\ }_{1V}c^{\ }_{1A}g^{\ }_{1V}g^{\ }_{1A}]
\nonumber\\
&\ &\hskip 1cm \times t(2s+t-M^2_L)\nonumber\\
&\ &\hskip .75cm + 4h_2[(c^2_{1V}+c^2_{1A})g^{\ }_{1V}g^{\
}_{1A}-(g^2_{1V}+g^2_{1A})
c^{\ }_{1V}c^{\ }_{1A}]\nonumber\\
&\ &\hskip 1cm \times s(s-M^2_L) \rbrace\nonumber\\
&-& 32\pi^2\alpha^2(1-h_1h_2)\lambda
{(s-M^2_Z)(t-M^2_B)+\Gamma_ZM_Z\Gamma_BM_B
 \over D(s,M_Z,\Gamma_Z)D(t,M_B,\Gamma_B)} \nonumber\\
&\ &\times\lbrace (1-h_1h_2)[(g^{\ }_{1V}g^{\ }_{2V}+ g^{\ }_{1A}
g^{\ }_{2A})(c^{\ }_{1V}c^{\ }_{2V}+c^{\ }_{1A}c^{\ }_{2A})\nonumber\\
&\ &\hskip 2.cm +(g^{\ }_{1V}g^{\ }_{2A}+g^{\ }_{1A}g^{\ }_{2V})
(c^{\ }_{1V}c^{\ }_{2A}+c^{\ }_{1A}c^{\ }_{2V})]\nonumber\\
&\ &\hskip 0.5cm + (h_2-h_1) [(g^{\ }_{1V}g^{\ }_{2V}+g^{\ }_{1A}g^{\  }_{2A})
(c^{\ }_{1V}c^{\ }_{2A}+c^{\ }_{1A}c^{\ }_{2V})\nonumber\\
&\ &\hskip 2.cm +(g^{\ }_{1V}g^{\ }_{2A}+g^{\ }_{1A}g^{\ }_{2V})
(c^{\ }_{1V}c^{\ }_{2V}+c^{\ }_{1A}c^{\ }_{2A})]
\rbrace\nonumber\\
&\ &\hskip 1.cm \times(s+t)(s+t-M^2_L)
\end{eqnarray}
with $D(q^2,m,\Gamma) \equiv (q^2-m^2)^2+(m\Gamma)^2$.
\par
In the equation above, $h_1(h_2)$ stands for the helicity of the incoming
electron (positron).  For a W(Z) boson exchange in the t-channel,
$\lambda=1(-1)$,
$M_B=M_W(M_Z)$ and ${\Gamma}_B=\Gamma_W(\Gamma_Z)$.   Vertices
$e^+ \bar{\nu} W^-$ and $e^+e^-Z^0$ are written as
$-ie{\gamma}_{\mu}(g_{kV}-g_{kA}{\gamma}_5)$, whereas the nonstandard vertices
$e^{\pm}L^-Z^0$ and $e^-W^+L^0$ are expressed as
$-ie{\gamma}_{\mu}(c_{kV}-c_{kA}{\gamma}_5)$, where $k=1$ refers to t-channel
$Z/W$ exchange and $k=2$ to the s-channel process.

\end{document}